# Water is not a Dynamic Polydisperse Branched Polymer

Teresa Head-Gordon[1-4*]
[1]Pitzer Theory Center, Departments of [2]Chemistry, [3]Bioengineering, [4]Chemical and Biomolecular Engineering, University of California, Berkeley,
Chemical Sciences Division, Lawrence Berkeley National Laboratory
Berkeley, CA 94720

Francesco Paesani[5-7]
[5]Department of Chemistry and Biochemistry, [6]Materials Science and Engineering, [7]San Diego Supercomputer Center, University of California, San Diego,
La Jolla, CA 92093

*corresponding author

The contributed paper by Naserifar and Goddard(1) reports that their RexPoN water model(2) under ambient conditions is comprised of a "dynamic polydisperse branched polymer", which they speculate explains the existence of the liquid-liquid critical point (LLCP) in the supercooled region. The observable they rely upon to support this is the oxygen–oxygen radial distribution function, $g_{OO}(r)$, from a dated neutron scattering experiment(1). Although it is well known that neutron scattering is almost exclusively sensitive to hydrogen correlations, and $g_{OO}(r)$ is more reliably obtained from X-ray scattering(3), they make the unsupported statement "the most reliable $g_{OO}(r)$ curve is neutron where there is no inference from electrons"(1). However, two X-Ray $g_{OO}(r)$ curves in Figure 1b of their paper are from a joint neutron-X-ray analysis (X-Ray1(4)) and neutron scattering study (X-Ray3(5)). Given the importance the authors have placed on $g_{OO}(r)$(1), it is evident that their RexPoN model is in disagreement with the most reliable estimate X-Ray 2 (3).

Naserifar and Goddard dismiss the MB-Pol model as inaccurate because the calculated coordination number ($N_c$) from $g_{OO}(r)$ is too high at 6.2(1). MB-Pol using path integral molecular dynamics(6) yields $N_c$=4.8, in good agreement with the experimental estimate $N_c$=4.5. Since RexPoN is fitted from quantum mechanics(1), the fact that reported properties are based on classical MD simulations(1) means that necessary inclusion of NQEs(7) will diminish their apparent agreement with experiment to yield a far less accurate model then claimed, especially for the heat of vaporization and liquid structure, the latter which is the lynchpin of their claims of superiority of RexPon.

Naserifar and Goddard claim that RexPoN water melts from four-coordinated ice into an ambient liquid with two strong hydrogen bonds that couple water molecules into isolatable regions resembling a branched nanopolymer. More specifically, the RexPon simulation melts from four-coordinated ice into an ambient liquid in which they observe that single hydrogen bonds (SHBs) "connect to form multibranched polymer chains (151 $H_2O$ per chain at 298 K), where branch points have 3 SHBs and termination points have 1 SHB". They then speculate using this "structural feature" for the existence of the LLCP at 227 K and 1 bar (1).

There are several serious factual errors and needless speculation in (1) on this point. Vibrational Raman data and X-ray spectroscopies taken on ambient water(8, 9) can be explained by a homogeneous liquid state without any unusual structural motifs.(10, 11) Even if a second critical point for supercooled water exists, it is irrelevant at 298 K since well above a critical point the state of matter is homogeneous and should not have isolatable structural motifs.(11, 12) Finally, recent work by Kim and co-workers have reported a *compressibility maximum* at 227 K and 1 bar(13), which is consistent with (but does not prove) a LLCP at likely a higher pressure and lower temperature. Well-validated models such as MB-Pol(6) and iAMOEBA(14) are reasonably accurate in the supercooled region for the structure factor and the isothermal compressibility to analyze experiments that claim to support the LLCP(13). However, these water models do not yield an ambient liquid with discernable structural domains, dynamic or not, but are more straightforwardly described as a deformable tetrahedral hydrogen-bonded network above a percolation threshold in which every water molecule experiences the same intermolecular force on average.

**Acknowledgments.** THG thanks the support from the Director, Office of Science, Office of Basic Energy Sciences, Chemical Sciences Division of the U.S. Department of Energy under Contract No. DE-AC02-05CH11231. FP thanks the National Science Foundation CHE-1453204.